\documentclass[twocolumn,preprintnumbers,amsmath,amssymb]{revtex4}
\usepackage{graphicx}
\usepackage{dcolumn}
\usepackage{bm}

\begin{document}


\title{Scaling of the fidelity susceptibility in a disordered quantum spin chain}

\author{N. Tobias Jacobson}
\email{ntj@usc.edu}
\affiliation{Department of Physics and Astronomy, University of Southern California, Los Angeles, CA 90089}

\author{Silvano Garnerone}
\email{garneron@usc.edu}
\affiliation{Department of Physics and Astronomy, University of Southern California, Los Angeles, CA 90089}
 
\author{Stephan Haas}
\affiliation{Department of Physics and Astronomy, University of Southern California, Los Angeles, CA 90089}

\author{Paolo Zanardi}
\altaffiliation[Also at ]{Institute for Scientific Interchange, Viale Settimio Severo 65, I-10133 Torino, Italy}
\affiliation{Department of Physics and Astronomy, University of Southern California, Los Angeles, CA 90089}

\date{\today}

\begin{abstract}
The phase diagram of a quantum XY spin chain with Gaussian-distributed random anisotropies and transverse fields is investigated, with focus on the 
fidelity susceptibility, a recently introduced quantum information theoretical measure.  Monitoring the finite-size scaling of the probability distribution of this 
quantity as well as its average and typical values, we detect a disorder-induced disappearance of criticality and the emergence of Griffiths phases in 
this model.  It is found that the fidelity susceptibility is not self-averaging near the disorder-free quantum critical lines.  At the Ising critical 
point the fidelity susceptibility scales as a disorder-strength independent stretched exponential of the system size, in contrast with the quadratic scaling at the corresponding point 
in the disorder-free XY chain.  Along the line where the average anisotropy vanishes the fidelity susceptibility appears to scale extensively, whereas in the disorder-free case 
this point is quantum critical with quadratic finite-size scaling.
\end{abstract}


\maketitle

\section{Introduction}
In the last few years, tools from the field of quantum information theory have found extensive use in the study of the phase 
diagrams of quantum systems.  One such technique, the fidelity approach to Quantum Phase Transitions (QPTs) has been 
successfully applied to various systems possessing quantum critical points \cite{PZ, ZhBa, ZhZhLi, Zh} (see 
\cite{Gu} for a review). This technique can be generalized to finite-temperature systems \cite{ZQWS,ZCVG}, 
classical phase transitions \cite{QC}, and topological phase transitions \cite{HaZhHa,AbZa,AbHaZa,YaGuSu,GaAbHa,Eriksson}.  

In a recent letter \cite{GJHZ} we have studied the fidelity in the context of 
disordered quantum systems. The physics peculiar to disordered quantum systems 
is reflected in the properties of the fidelity, a quantity not previously used to investigate such systems.
Here we study the scaling behavior and provide details concerning the zero-temperature phase diagram of the disordered quantum XY model in a transverse field, a 
prototypical model in the context of disordered quantum systems.

The paper is organized as follows. Section \ref{sec:met_mod} is devoted 
to defining the model, along with a review of known results about its phase diagram
and the basics of the fidelity approach. Section \ref{sec:res} presents 
the numerical results of our study and discusses the main features of the 
fidelity for disordered quantum chains. 
Our conclusions are presented in Section \ref{sec:conclusion}.

\section{Method and model}\label{sec:met_mod}
It is known that disorder can have interesting effects on a system's phase diagram \cite{IgMo}. 
In particular, Griffiths phases may arise as a result of the randomness \cite{Gri}.  
Here we study the disordered anisotropic quantum XY 
spin chain in a random transverse field, a model where the disorder-free case can be analytically solved 
\cite{LiScMa} and for which some exact results are known in the disordered case \cite{McK, BMcK}. Its Hamiltonian is given by
\begin{equation} \label{eq:XYHamiltonian}
H=-\sum_{i=1}^{L} 
\frac{1+\gamma_i}{2} \sigma_i^x \sigma_{i+1}^{x} +
\frac{1-\gamma_i}{2} \sigma_i^y \sigma_{i+1}^{y} + \lambda_i \sigma_i^z,
\end{equation}
where $\sigma_i^{\left\{x,y,z\right\}}$ are Pauli matrices, and the fields
$\lambda_i$ and anisotropies $\gamma_i$ are independent Gaussian-distributed random variables. 
The average field and anisotropy are denoted by $\lambda$ and $\gamma$, respectively. The variance is taken 
to be the same for both the field and anisotropy distributions.

The Jordan-Wigner transformation maps this system onto quasi-free spinless fermions \cite{LiScMa}.  
Neglecting the boundary term and taking the system to be closed in the fermion index, we obtain a Hamiltonian of the form
\begin{equation}\label{eq:Hamfree}
H=\sum_{i,j=1}^{L} c^{\dagger}_i A_{ij}c_j + 
\frac{1}{2} \sum_{i,j=1}^{L} \left( c^{\dagger}_i B_{ij}c^{\dagger}_j + c_j B_{ij} c_i\right),
\end{equation}
where $A$ and $B$ are symmetric and antisymmetric real $L \times L$ matrices, respectively. Explicitly:
$A_{ij} = -(2\lambda_i \delta_{ij} + \delta_{i,j+1} + \delta_{i+1,j}) $, $A_{1L} = A_{L1} = -1$ and 
$B_{ij} = \gamma_{j} \delta_{i,j+1} - \gamma_{i} \delta_{i+1,j}$, $B_{1L} = \gamma_{L} = -B_{L1}$.

The Hamiltonian may be 
rewritten in terms of the matrix $Z \equiv A - B$,  which contains all information about the 
system.  Performing the polar decomposition of $Z$ 
we obtain the matrices $\Lambda$ and $T$ 
such that $Z=\Lambda T$, where $\Lambda$ is a positive semi-definite 
matrix and 
$T$ is unitary. From the eigenvalues of $\Lambda$ one obtains the single-particle energy spectrum \cite{CGZ}.
  
For systems at zero temperature, the fidelity is simply the absolute value of the overlap 
between ground states corresponding to nearby points in parameter space. 
Near a quantum critical point the ground state changes rapidly 
for small shifts in the tuning parameters, an effect which is reflected in a corresponding decrease of the fidelity.

The ground state fidelity can be cast in terms of the unitary matrix $T$ in the following way \cite{ZCG}
\begin{equation}\label{eq:fid}
F(Z,\tilde{Z})=\sqrt{\vert \det{\frac{T+\tilde{T}}{2}} \vert},
\end{equation}
where $T$ and $\tilde{T}$ are respectively 
the unitary parts of the matrices $Z \equiv Z(x)$ and $\tilde{Z} \equiv Z(x')$,  
evaluated at the model parameters $ x $ and $ x' $. 
The corresponding fidelity susceptibility is defined as \cite{YouLiGu,CVZ}
\begin{equation}\label{eq:chi}
\chi (x)=\lim_{\Delta x \rightarrow 0} \frac{-2 \ln{F(x,x + \Delta x)}}{\Delta x^2}, 
\end{equation}
and can be written in terms of the unitary matrix $T$ as
\begin{equation}\label{eq:chifrob}
\chi (x) =\frac{1}{8}\Vert\partial_{x}T \Vert_{F}^{2},
\end{equation}
where $\Vert \cdot \Vert_{F}$ is the Frobenius norm. For a derivation of Eq. (\ref{eq:chifrob}) see the Appendix.  

We evaluate the fidelity susceptibility 
using (\ref{eq:chifrob}) by performing a singular value decomposition of $Z$
$$
Z= U \Sigma V^{\dagger} = (U \Sigma U^{\dagger}) (U V^{\dagger}) = \Lambda T,
$$ 
where $U$ and $V$ are unitary matrices.  
Note that the fidelity susceptibility is defined for infinitesimally separated points along any chosen direction 
in parameter space.

\subsection{Disorder-free case}
\begin{figure}[htp]
\includegraphics[scale=0.25]{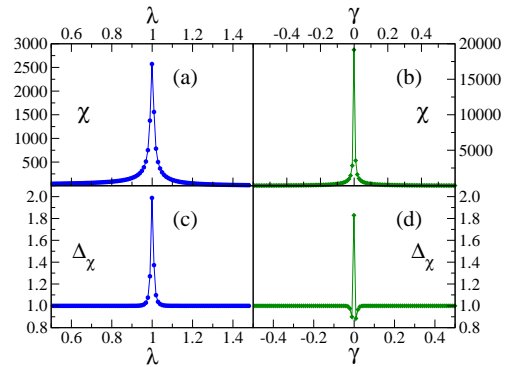} 
\caption{(color online) The disorder-free case: (a) Fidelity susceptibility near the Ising transition, with $\gamma = 1$ and system size $L=500$. (b) Near the anisotropy transition, 
with $\lambda = 0.5$ and system size $L=500$. (c) Finite-size scaling dimension of the fidelity susceptibility near the Ising transition, with $\gamma = 1$. The sizes considered range from 
$L=100$ to $700$, with $\chi \sim L^{\Delta_{\chi}}$. (d) Finite-size scaling of the fidelity susceptibility near the anisotropy transition, with $\lambda= 0.5$. The sizes considered range from $L=100$ to $600$.}
\label{fig:Fig1}
\end{figure}

Before considering the effects of disorder in the XY chain, let us first recall the behavior of the fidelity susceptibility for the 
disorder-free case, where 
$\lambda_i = \lambda, \gamma_i = \gamma, \forall i \in \{1,...,L\}$ \cite{PZ}.  The system can then be found in one of three phases.  For 
$|\lambda| > 1$ it is paramagnetic, and for $|\lambda| < 1$ and $\gamma > 0$ ($\gamma < 0$) the system is ferromagnetic along the $x$-
direction ($y$-direction). 
The boundary between any two of these phases is a quantum-critical line corresponding to a second-order quantum phase 
transition. 
Here, we refer to the transition driven by the magnetic field as the Ising transition, and to the transition 
driven by the anisotropy coupling as the anisotropy transition.
At the quantum-critical points there is an avoided level crossing between the ground state and the first excited state. 
As shown in Figs. \ref{fig:Fig1}(a) and \ref{fig:Fig1}(b), one observes
a maximum of the fidelity susceptibility at both the Ising and anisotropy critical lines. 

Moreover, the finite-size scaling dimension of the fidelity susceptiblity in Figs. \ref{fig:Fig1}(c) 
and \ref{fig:Fig1}(d) show it to be extensive away from criticality and superextensive (scaling quadratically 
with $L$) at the critical points. 
This scaling behavior holds for both the Ising and anisotropy critical lines. The apparent subextensive scaling in 
the immediate vicinity of the anisotropy critical point is a numerical artifact due to the narrowing of the fidelity susceptibility peak as the system size grows.  The 
Ising transition does not show this behavior, since the narrowing appears to occur more slowly than for the anisotropy transition.

It has been shown \cite{CVZ}, for translationally invariant systems, that superextensive finite-size scaling 
of the fidelity susceptibility implies a vanishing gap and therefore quantum criticality.  As a result, for clean 
systems the points in parameter space corresponding 
to superextensive scaling of this quantity mark quantum critical regions. When randomness is introduced, 
translational invariance is lost, and hence superextensive scaling of the fidelity susceptibility does not necessarily imply 
quantum criticality. As discussed before, we find that locations of superextensive scaling reveal more general behavior beyond quantum 
criticality, namely Griffiths phenomena \cite{Gri}.

\subsection{Random XY chain}
The effects of disorder on the physics of quantum magnets has 
been studied mainly using the strong-disorder renormalization 
group technique (SDRG) \cite{DasMa,Fis1, Fis2}. 
A different approach has been used in the work of McKenzie and Bunder \cite{McK, BMcK},
where the critical behavior of the 
disordered XY chain has been studied using 
a mapping to random-mass Dirac equations. 
The properties of the solutions of these equations imply the 
disappearance of the anisotropy transition in the presence of 
disorder. Furthermore, Griffiths phases are predicted to appear 
both around the Ising critical line and the anisotropy $\gamma=0$ line.
These results, together with the analysis performed by Fisher \cite{Fis1, Fis2}, 
are significant since they analytically show the drastic effects that disorder 
can have on the critical properties of a quantum system.

At fixed $\gamma$
the XY random chain is closely related to the random transverse-field Ising chain
(RTFIC), which is another prototypical model for disordered quantum systems \cite{Fis2}. 
Since the RTFIC is representative of the universality class of Ising transitions 
for all values of $ \gamma$, let us
review what is known for this model. 
The Hamiltonian of the RTFIC is
$H=-\sum_{i=0}^{L-1} 
\left[J_i \sigma_i^x \sigma_{i+1}^{x} +
h_i \sigma_i^z 
\right],
$
where $J_i$ and $h_i$ are random 
couplings and fields respectively. The system is critical when 
the average value of the field equals the 
average value of the coupling. 
Using the SDRG one obtains that, at the quantum critical 
point, the time scale $\tau$ and the length scale $L$ are related by 
$\ln{\tau} \sim L^{1/2}$. This results in an infinite 
value for the dynamical exponent $ z $ at criticality \cite{Fis2}. The distribution 
of the logarithm of the energy gap $\epsilon$ at criticality broadens
with increasing system size, in 
accordance with the scaling relation $\ln{\epsilon} \sim -L^{1/2}$ \cite{YoRi}. 
In the vicinity of the critical point 
the distribution of relaxation times is broad due to the presence of a Griffiths phase, 
characterized by a non-universal dynamical exponent $z$
depending on the distance from the critical point. This 
dependence can be used as an indicator for the Griffiths phase. 

\begin{figure}[htp]
\includegraphics[scale=0.25]{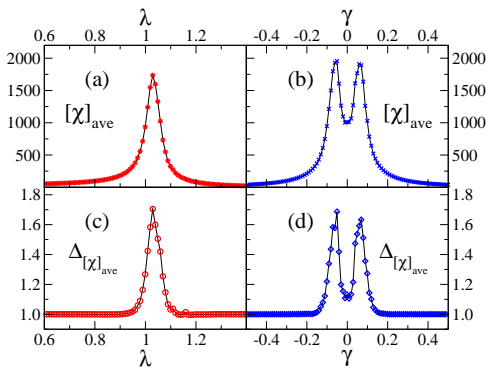} 
\caption{(color online) (a) Average fidelity susceptibility near the Ising transition, for $L=500$, $\gamma=1$, and $\sigma=0.3$, (b) Average $\chi$ near the $\gamma=0$ line, for $L=500$, $\lambda=0.5$, and $\sigma=0.3$, (c) Scaling dimension of $\chi$ for the same set of parameters near the Ising line, considering $L \in \{100,200,300,400,500\}$, (d) Scaling dimension of $\chi$ for the same set of parameters near the $\gamma=0$ line, considering $L \in \{100,200,300,400,500\}$.}
\label{fig:Fig2}
\end{figure}

In \cite{GJHZ} a study of the phase diagram of a random XY spin chain in a random transverse field was performed 
using the fidelity approach. There it was shown that superextensive finite-size scaling of $ \chi $ signals 
the presence of a quantum phase transition close to the Ising critical line, while 
the minimum in $ \chi $ close to the anisotropy $ \gamma =0 $ line 
is consistent with the absence of a phase transition in that parameter region, 
see Figs. \ref{fig:Fig2}(a) and (b).

The Griffiths phases of the model manifest themselves in a non-universal 
dependence of the finite-size scaling dimension $\Delta_{\chi}$ 
of the fidelity susceptibility on the distance from the disorder-free critical point, 
see Figs. \ref{fig:Fig2}(c) and (d). 
The relation between the dynamical scaling exponent $z$ 
and the scaling dimension of $ \chi $ \cite{CVZ}
\begin{equation}
\Delta_{\chi}=2 z + 2-2 \Delta_{O}
\label{eq:scaling_formula}
\end{equation} 
establishes the connection between the fidelity 
susceptibility and the Griffiths phase. In Eq.(\ref{eq:scaling_formula}) 
$\Delta_{O}$ is the scaling dimension of the relevant operator 
driving the transition. Eq. (\ref{eq:scaling_formula}) implies 
a non-universal scaling dimension for the fidelity susceptibility $ \Delta_{\chi} $, 
provided that the behavior of the 
unknown scaling dimension $\Delta_{O}$ does not exactly cancel that of the dynamical exponent. 
In the following we would like to study, using other methods, 
the extent of the Griffiths phase for this model.  In particular, we would like to verify 
that the entrance into a Griffiths phase is indeed reflected by a changing scaling behavior of the fidelity susceptibility.

In our numerical analysis we consider system sizes 
$L \in \{100, 200, 300, 400, 500\}$
and for each system size we compute $10^4$ disorder realizations.  
We take the external fields (anisotropies) to be independent 
and identically distributed Gaussian random variables with 
standard deviation $\sigma$ and mean $\lambda$ ($\gamma$).
We consider the range of values $\{0.1,0.2,0.3,0.4\}$ for the standard deviation $\sigma$. 
This disorder strength can be considered strong with respect to the value of the other parameters.
We denote with $\left[ \cdot \right]_{\textrm{ave}}$ the arithmetic mean over all $10^4$ disorder realizations.

The width of the Griffiths phase for the XY model with weak Gaussian disorder in the continuum limit is known due to 
the work of McKenzie \cite{McK}. Let us denote the distance from 
criticality with $\delta$, where $\delta=0$ 
corresponds to the points in parameter space where the pure system is critical. 
From \cite{McK} one can compute that 
near the Ising transition, where the field $\lambda$ drives the transition, 
$\delta = \frac{|\gamma| (\lambda - 1)}{\sigma^2}$, while 
near the anisotropy line $\delta = \frac{\gamma (1 - \lambda^2)}
{\sigma^2}$.

For the so-called commensurate case, which corresponds to the Ising transition for this system since we have disorder in 
both the field and anisotropy, McKenzie showed that the disorder-averaged density of states $\left[ \rho(E) \right]_{\rm{ave}} / \rho_0$ 
diverges at zero energy within the range $|\delta| < 1/2$ away from the critical point \cite{McK}. Here $\rho_0$ is the 
``high-energy" density of states.  This divergence implies that the gap distribution function $P(\Delta E)$ also diverges at 
zero energy, since a large density of states means a vanishingly small gap.  Note that a diverging probability of having a 
vanishing gap does not necessarily imply that the density of states is also divergent, since the gap distribution only provides information about the position of the first 
excited energy level relative to the ground state energy.  However, since a diverging gap distribution should be expected to 
occur as a result of a divergence of the low-energy density of states, we will use it to give a rough estimate of the extent of the Griffiths phase.

For the incommensurate case, which includes the anisotropy transition, $\left[ \rho(E=0) \right]_{\rm{ave}} / \rho_0$ is of the order of unity 
for some range of parameters about $\delta = 0$, implying effective gaplessness, and $\left[ \rho(E=0) \right]_{\rm{ave}} / \rho_0$ is much smaller 
than unity for $|\delta| >> 1$, giving an effectively finite gap \cite{McK}.  Note that the boundary of the Griffiths phase in this region is expected 
to be less defined than near the Ising transition, since the zero-energy density of states does not diverge at any value of $\delta$.

These results apply for the case of weak disorder, but we consider a range of moderate to strong disorder 
strengths.  In order to compare with the results of McKenzie for the Griffiths phase extent, we propose a rough 
criterion for determining the extent of the Griffiths phase using the gap distribution.  Assume that the Griffiths 
phase lies within the range of parameter values for which the distribution $P(\Delta E)$ has a maximum for $\Delta E 
= 0$.  At some point away from criticality 
the distribution maximum moves away from zero, eventually becoming approximately Gaussian 
far from the disorder-free critical point. We have determined the range of parameters for which 
$P(\Delta E)$ has a maximum at zero gap.  The width of this range of parameters is independent of system size and 
scales with the variance of the coupling or field distributions, in accordance with the result of 
\cite{McK}.  However, this criterion gives an extent several times larger than the McKenzie value, around both 
the Ising and $\gamma=0$ lines.  
This difference may be due to strong disorder or, rather, the result 
of an overestimation of the Griffiths phase extent since the distribution $P(\Delta E)$ having a maximum at zero does 
not directly imply a divergent zero-energy density of states.  Nonetheless, consistency between 
different estimations of the Griffiths phase extent support the validity, at least qualitatively, of this approach.

\begin{figure}[htp]
\includegraphics[scale=0.25]{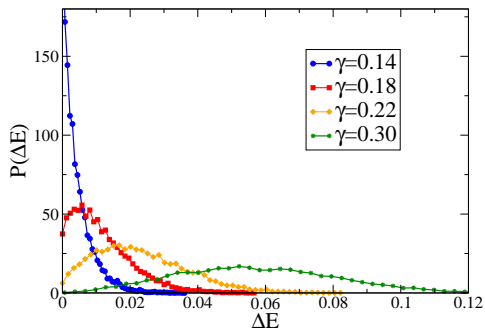} 
\caption{(color online) Gap distribution near the anisotropy line, with $\lambda=0.5$, $\sigma=0.3$, and system size $L=500$. Plotted 
distributions are for $\gamma = 0.14, 0.18, 0.22,$ and $0.30$.}
\label{fig:Fig3}
\end{figure}

Fig.(\ref{fig:Fig3}) shows the gap distribution in the vicinity of the $\gamma=0$ line.  For small 
anisotropies the distribution has a pronounced peak at zero gap.  Moving away from the anisotropy line, the distribution develops a peak 
slightly away from $\Delta E = 0$, and for larger anisotropies the distribution becomes Gaussian with a vanishing probability of having zero gap.  
Similar behavior holds for the Ising transition as the average field $\lambda$ is adjusted away from the finite-size pseudocritical point. \\

\section{Results}\label{sec:res}

\subsection{Average and typical values}
In computing the \emph{average} of some physical quantity, namely the arithmetic mean over many 
disorder realizations, any 
rare but large values will significantly affect the result.  On the other hand, the geometric mean over disorder 
realizations gives a more representative 
measure of the \emph{typical} values of the physical quantity.  Recall that the arithmetic mean provides an upper bound
for the geometric mean when averaging over 
a set of positive values, and the two are equal only when taking the mean of a constant set of values.

\begin{figure}[htp]
\includegraphics[scale=0.25]{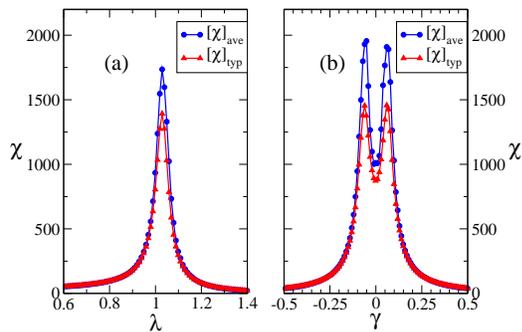} 
\caption{(color online) Average (blue) and typical (red) fidelity susceptibility about: (a) the Ising line, for $\gamma=1$, $L=500$, and $\sigma=0.3$ 
(b) the $\gamma=0$ line, for $\lambda=0.5$, $L=500$, and $\sigma=0.3$.}
\label{fig:Fig4}
\end{figure}

To observe the presence of large fluctuations in the fidelity susceptibility, we plot in Fig. \ref{fig:Fig4}(a) 
the disorder-averaged as well as typical fidelity susceptibility in the vicinity of the
Ising critical point for a disorder strength $\sigma=0.3$.  Notice that in the vicinity of the critical point the 
average becomes significantly 
larger than the typical value, indicating that there are instances of large fidelity susceptibilities that skew the 
arithmetic average towards a greater value. 

Near the anisotropy line, as shown in Fig. \ref{fig:Fig4}(b), there are also regions where the average 
fidelity susceptibility becomes 
much larger than the typical value, but now the positions of largest difference do not correspond to a critical 
point.  Indeed, at the point $\gamma=0$, 
which in the disorder-free case is critical, the average-typical difference is much smaller than it is at the two 
offset peaks. This is evidence for the disappearance of the anisotropy transition as a result of the disorder.

\begin{figure}[htp]
\includegraphics[scale=0.25]{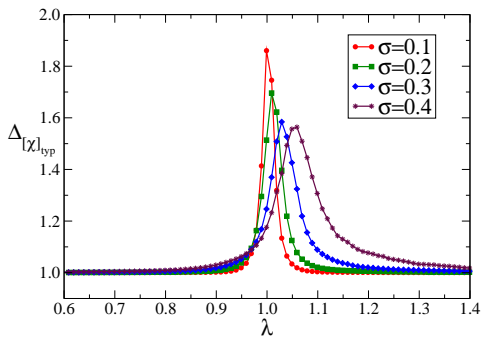} 
\caption{(color online) Finite-size scaling dimension of the typical fidelity susceptibility about the Ising transition, for $\gamma=1$, for $\sigma \in \{0.1,0.2,0.3,0.4\}$.  Scaling fit considered system sizes $L = 200,300,400,500$, with $\chi \sim L^{\Delta_{\left[\chi\right]_{typ}}}$ assumed.}
\label{fig:Fig5}
\end{figure}

For translationally invariant systems, which are disorder-free, it has been shown that superextensive 
finite-size scaling 
of the fidelity susceptibility implies quantum criticality \cite{CVZ}.  
The disordered XY chain does not have translational invariance, so locations 
of superextensive scaling do not necessarily imply criticality.  However, it is still useful to consider 
finite-size scaling, since comparison 
with the disorder-free case may suggest in what way the phase diagram changes as a result of disorder. 
Fig.(\ref{fig:Fig5}) shows the finite-size scaling dimension of the typical fidelity susceptibility near 
the Ising transition. The locations 
of the maxima of the typical fidelity susceptibility and finite-size scaling dimension coincide, and are shifted from 
the pure pseudocritical point due to finite-size effects.  Also, the maximum scaling dimension obtained with disorder is 
smaller than the pure case of quadratic scaling in $L$, and this maximum value decreases with increased disorder 
strength.\\

\begin{figure}[htp]
\includegraphics[scale=0.25]{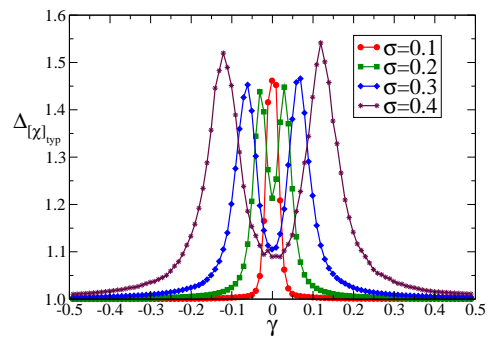} 
\caption{(color online) Finite-size scaling dimension of the typical fidelity susceptibility about the anisotropy line, for $\lambda=0.5$, for $\sigma \in \{0.1,0.2,0.3,0.4\}$.  Scaling fit considered system sizes $L = 100,200,300,400,500$, with $\chi \sim L^{\Delta_{\left[\chi\right]_{typ}}}$ assumed.}
\label{fig:Fig6}
\end{figure}

In Fig.(\ref{fig:Fig6}), the finite-size scaling dimension of $\chi$ around 
the anisotropy line $\gamma=0$ 
is shown for the same set of disorder strengths. 
Notice that the scaling depends on distance from the $\gamma=0$ line and at 
$\gamma=0$ the scaling is approximately 
extensive, as it is when far from the anisotropy line.  For sufficiently small disorder and system size there may 
appear to be only a single 
peak in the typical fidelity susceptibility at $\gamma=0$, suggesting that for these finite systems emergent criticality is 
still felt even though the 
quantum phase transition in the thermodynamic limit disappears as a result of the disorder.  However, increasing the 
system size reveals a double peak with a peak offset which grows with the strength of the disorder. In both the Ising 
and anisotropy regions, the width of the parameter interval giving scaling dimensions larger than a particular 
value scales approximately with the variance of the disorder distribution.  This scaling behavior agrees with that given by the  
previously mentioned gap distribution criterion for the Griffiths phase.

\begin{figure}[htp]
\includegraphics[scale=0.25]{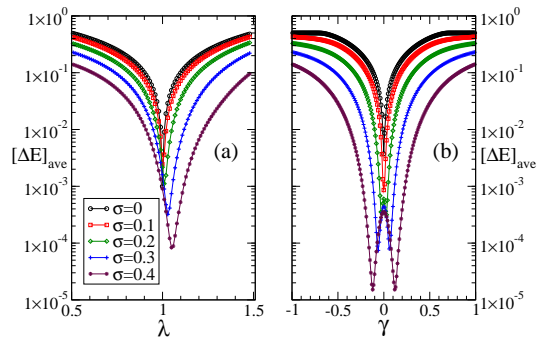} 
\caption{(color online) (a) Disorder-averaged gap near the Ising transition for disorder strengths $\sigma \in \{0, 0.1, 0.2, 0.3, 0.4\}$, $L=500$, and $\gamma=1$. (b) Disorder-averaged gap near the $\gamma=0$ line for the same range of disorder strengths, $L=500$, and $\lambda=0.5$.}
\label{fig:Fig7}
\end{figure}

In Fig.(\ref{fig:Fig7}) 
the disorder-averaged gap is plotted for the four disorder strengths.  The behavior 
of the gap corresponds closely with that of the fidelity susceptibility, 
in that the average gap minima have the same location as 
the typical $\chi$ maxima.  Note that the vicinity of the Ising critical line and the $\gamma=0$ line are regions of effective gaplessness which we 
associate with quantum criticality or Griffiths phases. \\

\begin{figure}[htp]
\includegraphics[scale=0.3]{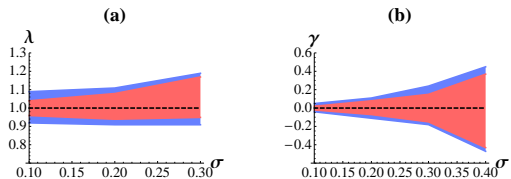} 
\caption{(color online) Regions about the Ising transition (a) and anisotropy line (b) for which $\chi$ is weakly self-averaging (blue) and non-self-averaging (orange), as a function of disorder strength $\sigma$.  Outside these regions $\chi$ is self-averaging.}
\label{fig:Fig8}
\end{figure}

An experiment on such a disordered system would consider only one particular realization of disorder, and as a result 
would not necessarily observe the disorder-averaged value of an observable but rather a typical value.  Here, we would like to study 
whether a measurement of the fidelity susceptibility for a large system coincides with the average value, 
and to do this we must see for what conditions 
$\chi$ is a \emph{self-averaging} quantity \cite{AH}.  Consider the quantity $R_{\chi}(x,L) = \textrm{Var}
[\chi(x)]/[\chi(x)]_{\rm{ave}}^2$.  We expect that 
$R_{\chi}(x,L)$ 
for fixed $x$ will scale as a power law in the system size $L$, $R_{\chi} \sim L^b$.  If $b=-1$ then we say $\chi$ is 
self-averaging, if $b<0$ then $\chi$ is weakly 
self-averaging, and if $b > 0$ then $\chi$ is not self-averaging.  In Fig.(\ref{fig:Fig8}) 
we indicate the 
regions for which $\chi$ is self-averaging, weakly self-averaging and non-self-averaging for various disorder 
strengths near the Ising transition as well as the $\gamma=0$ line.

\subsection{Fidelity susceptibility distributions}
\subsubsection{Near the Ising line}
\begin{figure}[htp]
\includegraphics[scale=0.25]{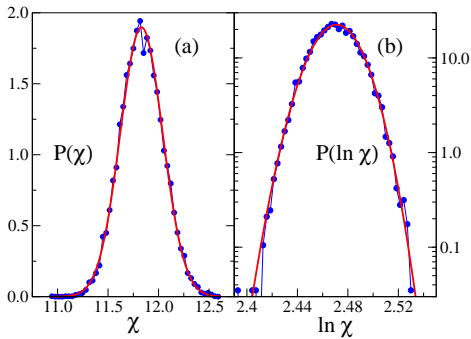} 
\caption{(color online) (a) Distribution of $\chi$ far from the Ising critical point, with $\lambda=1.49$ (blue). Overlaid is a Gaussian distribution function with the same mean and variance (red). (b) Distribution of $\ln \chi$ for the same parameters (blue) along with the corresponding distribution function of the logarithm of a Gaussian random variable.  Here $\sigma=0.1$, $L=500$ and $\gamma=1$. This distribution is well described as Gaussian.}
\label{fig:Fig9}
\end{figure}

\begin{figure}[htp]
\includegraphics[scale=0.25]{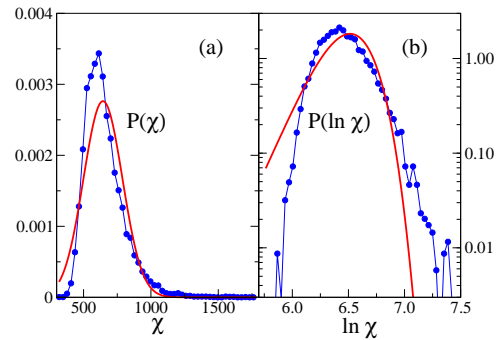} 
\caption{(color online) (a) Distribution near the Ising critical point, with $\lambda=1.03$ (blue). Overlaid is a Gaussian distribution function with the same mean and variance (red). (b) Distribution of $\ln \chi$ for the same parameters (blue) along with the corresponding distribution function of the logarithm of a Gaussian 
random variable.  Here $\sigma=0.1$, $L=500$, and $\gamma=1$.  Clearly this distribution is not Gaussian, as can be seen by the much slower 
dropoff of the tail towards large fidelity susceptibilities.}
\label{fig:Fig10}
\end{figure}

Far from the Ising critical line, the distribution of the fidelity susceptibility is  Gaussian, 
see Fig.(\ref{fig:Fig9}).  However, in the vicinity of the Griffiths phase and the 
critical point the distribution is non-Gaussian, developing a slowly-decaying tail towards large fidelity 
susceptibilities, as shown in Fig.(\ref{fig:Fig10}).  This tail reflects the presence of rare but large fidelity susceptibilities, and 
is expected to arise either in a Griffiths phase or in a quantum critical region.
\\

\begin{figure}[htp]
\includegraphics[scale=0.25]{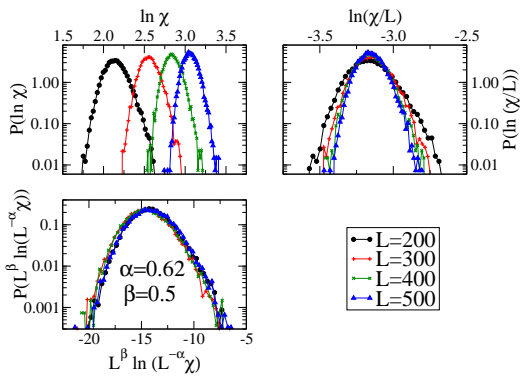} 
\caption{(color online) (a) Distribution $P(\ln \chi))$ at $\lambda=1.39$ for $L \in \{200,300,400,500\}$. (b) Distribution $P(\ln (\chi / L))$. (c) Rescaled distribution, $L^{\beta} \ln (L^{-\alpha} \chi)$.  Here $\sigma=0.2$, $\gamma=1$, and the fit parameters are $\alpha=0.62$ and $\beta=0.5$. Including this exponential correction improves the collapse of the curves. This scaling would suggest $\chi \sim L^{\alpha} \exp{L^{-\beta}}$.}
\label{fig:Fig11}
\end{figure}

Now we explore how the distribution of the logarithm of the fidelity susceptibility changes as the system size is varied, with all other 
parameters fixed.  Far from the Ising transition the distribution of $\ln(\chi)$ narrows with increasing system size, 
as shown in Fig. \ref{fig:Fig11}(a).  The position of the peak of the distribution remains fixed for a 
rescaling $\chi \to \chi/L$ (see Fig. \ref{fig:Fig11}(b)), but the width of the distribution decreases 
slightly.  Choosing a more general scaling assumption $\ln(\chi) \to L^{\beta} \ln (\chi / L^{\alpha})$ allows for an improved collapse 
of the distributions in this region, indicated in Fig. \ref{fig:Fig11}(c). The fit parameter $\alpha$ 
essentially translates the distribution, while the parameter $\beta$ adjusts the width.  Such a scaling would imply a size dependence 
$\chi \sim L^{\alpha} \exp{L^{-\beta}}$, where $\alpha, \beta \geq 0$.  However, for asymptotically large system sizes this would lead to subextensive 
scaling, and would thus not be expected to hold for all $L$. It appears that this apparent non-power-law scaling for the 
range of sizes we have considered may be due to finite-size effects, and we speculate that an assumption 
of extensive scaling would lead to an improved collapse for sufficiently large system sizes.

\begin{figure}[htp]
\includegraphics[scale=0.25]{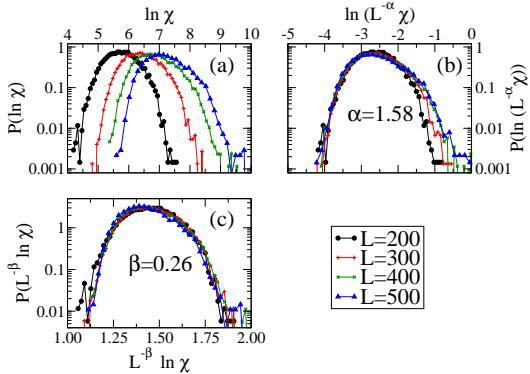} 
\caption{(color online) (a) Distribution $P(\ln \chi))$ at the pseudocritical point $\lambda=1.02$ for $L \in \{200,300,400,500\}$. (b) Distribution after a power-law rescaling $\ln (L^{-\alpha} \chi)$, where $\alpha=1.58$. (c) Distribution of $L^{-\beta} \ln \chi$ for $\beta=0.26$ and the same parameters.  Here $\sigma=0.2$ and $\gamma=1$, though this collapse applies for all other disorder strengths considered.}
\label{fig:Fig12}
\end{figure}

At the Ising pseudocritical point the distribution $P(\ln \chi)$ broadens significantly with increasing system size, 
and a rescaling $\ln(\chi) \to L^{-\beta} \ln(\chi)$ gives a good collapse for a value of the fit parameter $\beta=0.26$ (see 
Fig.(\ref{fig:Fig12})).  This collapse and value of fit parameter holds for the pseudocritical points corresponding 
to all four disorder cases we have considered.  A rescaling of this kind suggests that 
the fidelity susceptibility scales as a stretched exponential of the system size at the critical point rather than 
quadratically as in the pure XY chain. 
For the random transverse field Ising chain \cite{YoRi}, it is known that the energy gap vanishes as $\Delta E \sim 
\exp{-\sqrt{L}}$ at the Ising critical point.   Recalling the alternative expression for the fidelity susceptibility 
$\chi = \sum_{n \neq 0}{} \frac{|<n|\partial_{x}H|0>|^2}{|E_n - E_0|^2}$ \cite{YouLiGu}, we expect that the first 
term in this series would dominate, and that the fidelity susceptibility might scale as $\chi \sim 1 / (\Delta E)^2$.  
However, this crude argument appears not to be consistent with the scaling of the energy gap of the RTFIM, perhaps 
because of a lack of universality in the power of $L$ in the stretched exponential.

\subsubsection{Near the anisotropy line}
\begin{figure}[htp]
\includegraphics[scale=0.25]{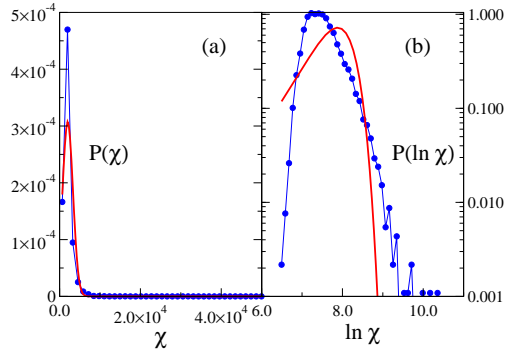} 
\caption{(color online) (a) Distribution of $\chi$ at the anisotropy line $\gamma=0$ (blue). Overlaid is a Gaussian distribution function with the same mean and variance (red). (b) Distribution of $\ln \chi$ for the same parameters (blue) along with the corresponding distribution function of the logarithm of a Gaussian 
random variable.  Here $\sigma=0.3$, $L=500$, and $\lambda=0.5$.  The distribution is non-Gaussian with a power-law tail towards large fidelity susceptibilities.}
\label{fig:Fig13}
\end{figure}

\begin{figure}[htp]
\includegraphics[scale=0.25]{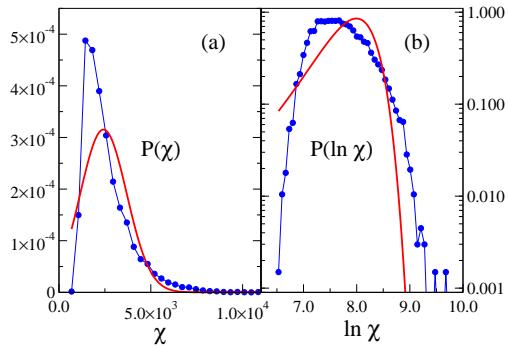} 
\caption{(color online) (a) Distribution for $\gamma=0.03$ (near the value of $\gamma$ coinciding with the peak in $\left[\chi\right]_{typ}$) (blue). Overlaid is a Gaussian distribution function with the same mean and variance (red). (b) Distribution of $\ln \chi$ for the same parameters (blue) along with the corresponding distribution function of the logarithm of a Gaussian random variable.  Here $\sigma=0.3$, $L=500$, and $\lambda=0.5$. Note that there does not appear to be a power-law tail towards large fidelity susceptibilities.}
\label{fig:Fig14}
\end{figure}

Just like for the Ising case, far from the anisotropy line the distribution $P(\chi)$ is well-approximated as a 
Gaussian. Closer to the $\gamma=0$ line the distribution looks much like the distribution of $\ln \chi(\lambda)$ 
in the vicinity of the Ising line, becoming non-Gaussian with a slowly-decaying tail towards large values 
of $\chi$. Fig.(\ref{fig:Fig13}) shows the distribution $P(\ln(\chi))$ for $\gamma=0$ and noise strength 
$\sigma=0.3$, and Fig.(\ref{fig:Fig14}) shows the same quantity for $\gamma=0.03$, the value of average 
anisotropy coinciding with the peak in the typical value of $\chi$ for that magnitude of disorder.

\begin{figure}[htp]
\includegraphics[scale=0.25]{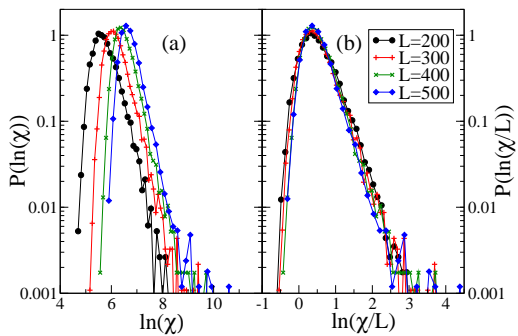} 
\caption{(color online) (a) Distribution $P(\ln \chi))$ at $\gamma=0$ for $L \in \{200,300,400,500\}$. (b) Rescaled distribution, $\ln (\chi / L)$ for the same parameters.  Here $\sigma=0.3$ and $\lambda=0.5$, though this collapse applies for all other disorder strengths considered.}
\label{fig:Fig15}
\end{figure}

\begin{figure}[htp]
\includegraphics[scale=0.25]{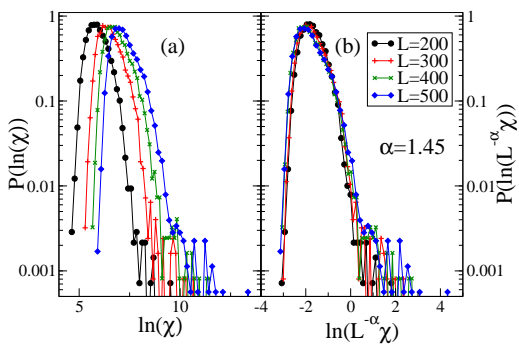} 
\caption{(color online) (a) Distribution $P(\ln \chi))$ at $\gamma=0.06$, the position of the maximum of $\left[\chi\right]_{typ}$ for $\sigma=0.3$. System sizes $L \in \{200,300,400,500\}$ are plotted. (b) Rescaled distribution, $\ln (L^{-\alpha} \chi)$ for the same parameters, with $\alpha=1.45$.  Here $\sigma=0.3$ and $\lambda=0.5$.}
\label{fig:Fig16}
\end{figure}

Considering the point $\gamma=0$, as the system size increases the distribution $P(\ln \chi)$ does not change width, so a rescaling 
$\chi \to \chi / L$ gives a good collapse, see Fig.(\ref{fig:Fig15}).  This scaling also agrees with the extensive 
scaling of the average fidelity susceptibility at $\gamma=0$.  Moving $\gamma$ away from this point in 
either direction, soon the distribution begins to shift superextensively, as shown in Fig.(\ref{fig:Fig16}).  For 
all values of $\gamma$ in this peak region, a rescaling of the form $\chi \to \chi / L^{\alpha}$ gives a good collapse, 
where $\alpha$ is the fit value of the finite-size scaling dimension of the corresponding typical fidelity susceptibility. 
Continuing to move $\gamma$ away from the peak in the typical fidelity susceptibility, the distribution begins to narrow 
slightly as in the off-critical Ising case.  However, a rescaling $\chi \to \chi / L$ appears to give a good collapse 
for $\gamma$ sufficiently large in magnitude.
\\

\section{Conclusion}\label{sec:conclusion}
In this work we have studied the effect of random transverse fields and couplings on the phase diagram of the quantum XY chain.  
By examining the finite-size scaling of the typical fidelity susceptibility and the fidelity susceptibility 
distribution for a range of disorder strengths and system sizes, we find agreement with earlier analytic results pertaining to the limit of weak 
randomness. The introduction of disorder clearly removes the anisotropy quantum critical line, replacing it with an extended Griffiths 
phase.  There, the typical fidelity susceptibility's finite-size scaling dimension 
depends strongly on the average value of the anisotropy parameter, and appears to become extensive at the line of vanishing anisotropy.  
At the Ising critical line, the stretched exponential scaling of the fidelity susceptibility distribution is consistent with what is expected of an infinite randomness fixed point, while a Griffiths phase is observed to form in the vicinity.
Remarkably, 
the scaling of the fidelity susceptibility distribution at the Ising critical line is universal, in that all disorder strengths give the same scaling behavior. In the Griffiths phase 
the fidelity susceptibility is not self-averaging.  However, self-averaging behavior returns sufficiently far from the disorder-free critical lines.  These detailed results suggest that the fidelity susceptibility may be a useful tool for the study of other disordered systems.

We would like to thank N. Bray-Ali for helpful comments. 
Computation for the work described in this paper was supported by the 
University of Southern California Center for High Performance Computing 
and Communications. We acknowledge financial support by the National 
Science Foundation under grant DMR-0804914.

\section{Appendix}

Here we derive an expression for the fidelity susceptibility 
$\chi$ in terms of the unitary matrix $T$:
\begin{eqnarray}\label{eq:derivation_1}
\lefteqn{F(Z,\tilde{Z})=\sqrt{\vert \det{\frac{T+\tilde{T}}{2}} \vert}}\nonumber \\
&=&\exp\left\{\textrm{Tr} \ln{\left(\frac{1+T^{\dagger}\tilde{T}}{2}\right)^{1/2}}\right\}\nonumber \\
&=&\exp\left\{\textrm{Tr} \ln{\left(\frac{1+T^{\dagger}\left(T+\delta T \right) }{2}\right)^{1/2}}\right\}\nonumber \\
&=&\exp\left\{\textrm{Tr} \ln{\left(1+\frac{T^{\dagger}\delta T}{2}\right)^{1/2}}\right\}\nonumber \\
&=&\exp\left\{\textrm{Tr} \frac{1}{2}\ln{\left(1+\frac{T^{\dagger}\delta T}{2}\right)}\right\}\nonumber \\
&\approx&\exp\left\{\textrm{Tr} \frac{1}{2} \left[ \frac{1}{2}T^{\dagger}\delta T-\frac{1}{8}\left( T^{\dagger}\delta T \right)^2 \right]\right\},
\end{eqnarray}
where $\delta T=\partial_{x}T d x $.  This leads to
\begin{eqnarray}\label{eq:derivation_2}
\lefteqn{F(Z,\tilde{Z})=\exp\left\{\textrm{Tr} \frac{1}{2}\left[\frac{1}{4}T^{\dagger}\partial_{x}^{2}Td x^{2}-\frac{1}{8}T^{\dagger}\partial_{x}T T^{\dagger}\partial_{x}Tdx^{2}  \right]\right\}}\nonumber \\
&=&\exp\left\{-\frac{1}{8}\Vert\partial_{x}T\Vert_{F}^{2}dx^2-\frac{1}{16}\textrm{Tr} \left[ T^{\dagger}\partial_{x}TT^{\dagger}\partial_{x}T \right]dx^2\right\}\nonumber \\
&=&\exp\left\{-\frac{1}{8}\Vert\partial_{x}T\Vert_{F}^{2}dx^2-\frac{1}{16}\textrm{Tr} \left[ -\partial_{x}T^{\dagger}TT^{\dagger}\partial_{x}T \right]dx^2\right\}\nonumber \\
&=&\exp\left\{-\frac{1}{8}\Vert\partial_{x}T\Vert_{F}^{2}dx^2-\frac{1}{16}\textrm{Tr} \left[ -\partial_{x}T^{\dagger}\partial_{x}T \right]dx^2\right\}\nonumber \\
&=&\exp\left\{-\frac{1}{16}\Vert\partial_{x}T \Vert_{F}^{2}dx^{2}\right\},
\end{eqnarray}
where we have used the anti-symmetry of $T^{\dagger}\partial_{x}T$ which implies 
$\textrm{Tr}\left[ T^{-1}\partial_{x}^{2}T \right]=-\Vert\partial_{x}T  \Vert_{F}^2$, with 
$\Vert \cdot \Vert_{F}$ the Frobenius norm. 
From (\ref{eq:chi}) it follows that
\begin{equation}\label{eq:chifrobappendix}
\chi=\frac{1}{8}\Vert\partial_{x}T \Vert_{F}^{2}.
\end{equation}

\end{document}